\documentclass[letterpaper]{article} 
\usepackage{aaai2026}  
\usepackage{times}  
\usepackage{helvet}  
\usepackage{courier}  
\usepackage[hyphens]{url}  
\usepackage{graphicx} 
\usepackage{natbib}  
\usepackage{caption} 
\usepackage{algorithm}
\usepackage{algorithmic}

\usepackage[centertags]{amsmath}
\usepackage{amsfonts}
\usepackage{booktabs}
\usepackage[table]{xcolor} 
\usepackage{amssymb}
\usepackage{multirow}
\usepackage{xcolor}
\usepackage{tabularray}
\UseTblrLibrary{booktabs}
\usepackage{newfloat}
\usepackage{listings}
\DeclareCaptionStyle{ruled}{labelfont=normalfont,labelsep=colon,strut=off} 
\floatstyle{ruled}
\newfloat{listing}{tb}{lst}{}
\floatname{listing}{Listing}
\title{Augmented Runtime Collaboration for Self-Organizing Multi-Agent Systems: \\A Hybrid Bi-Criteria Routing Approach}
\author {
    Qingwen Yang\textsuperscript{\rm 1}\equalcontrib\thanks{Author Contributions: Yang led the creation of the GapScore dataset and was responsible for training BiRouter.}, 
    Feiyu Qu\textsuperscript{\rm 2}\equalcontrib\thanks{Author Contributions: Qu designed the overall network framework and the BiRouter forwarding mechanism, and contributed to building the ImpScore dataset. Qu and Yang jointly executed all experiments  and collaborated on writing the manuscript.}, 
    Tiezheng Guo\textsuperscript{\rm 1}, 
    Yanyi Liu\textsuperscript{\rm 1}, 
    Yingyou Wen\textsuperscript{\rm 1}
}
\affiliations {
    \textsuperscript{\rm 1}School of Computer Science and Engineering, Northeastern University, Shenyang 110819, China\\
    \textsuperscript{\rm 2}Department of Computer Science, Dartmouth College, Hanover, NH 03755, USA\\
    2290182@stu.neu.edu.cn, feiyu.qu.gr@dartmouth.edu, 2310725@stu.neu.edu.cn, 2290175@stu.neu.edu.cn, wenyingyou@mail.neu.edu.cn
}

\usepackage{bibentry}

\begin{document}

\maketitle

\begin{abstract}
LLM-based multi-agent systems have demonstrated significant capabilities across diverse domains. However, the task performance and efficiency are fundamentally constrained by their collaboration strategies. Prevailing approaches rely on static topologies and centralized global planning, a paradigm that limits their scalability and adaptability in open, decentralized networks. Effective collaboration planning in distributed systems using only local information thus remains a formidable challenge. To address this, we propose BiRouter, a novel dual-criteria routing method for Self-Organizing Multi-Agent Systems (SO-MAS). This method enables each agent to autonomously execute ``next-hop'' task routing at runtime, relying solely on local information.  Its core decision-making mechanism is predicated on balancing two metrics: (1) the ImpScore, which evaluates a candidate agent's long-term importance to the overall goal, and (2) the GapScore, which assesses its contextual continuity for the current task state. Furthermore, we introduce a dynamically updated reputation mechanism to bolster system robustness in untrustworthy environments and have developed a large-scale, cross-domain dataset, comprising thousands of annotated task-routing paths, to enhance the model's generalization. Extensive experiments demonstrate that BiRouter achieves superior performance and token efficiency over existing baselines, while maintaining strong robustness and effectiveness in information-limited, decentralized, and untrustworthy settings.
\end{abstract}


\section{Introduction}

Recent breakthroughs in Large Language Models (LLMs) such as GPT-4 \cite{GPT4} have unlocked unprecedented capabilities in autonomous reasoning and planning \cite{CoT,SC-CoT,ReAct,LLM-survey1}. When embodied as agents \cite{ModelScopeAgent,AutoFlow}, these models can comprehend complex instructions, decompose tasks, and execute actions to solve problems in diverse domains \cite{Agent-survey,Agent-survey1}. This success has catalyzed a shift from single-agent systems to the more ambitious paradigm of Multi-Agent Systems (MAS) \cite{chateval,MAS-survey}. By orchestrating agents with distinct roles and specialized skills, MAS can achieve a sophisticated division of labor, mirroring human societal structures to tackle tasks of far greater complexity \cite{MAC-survey}. However, orchestrating this collaboration, especially in open and decentralized environments, remains a formidable scientific challenge.

The prevailing paradigm for MAS development has been the closed-world system, which relies on global optimization to coordinate agents \cite{AgentPrune}. This approach inherently limits scalability and adaptability, prompting a community-wide effort towards open agent networks that integrate heterogeneous agents from different systems. Foundational protocols such as Agent-to-Agent (A2A) \cite{A2A} and Agent Network Protocol (ANP) \cite{ANP} have emerged from this effort. However, these protocols provide mechanisms for agent discovery and invocation, but lack the strategies for dynamic, task-oriented collaboration. This absence leads to sub-optimal, or even chaotic, agent interactions, resulting in wasted computational resources, cascading failures, and an inability to reliably solve complex, multi-step tasks. Existing planning methods like GPTSwarm \cite{GPTSwarm} and G-Designer \cite{G-Designer} rely on static, global plans that pre-compute a complete execution path. Such an approach is not only brittle to dynamic changes but fundamentally incompatible with decentralized settings where no single agent possesses a global view. Furthermore, their generalization capabilities are often limited as open environments demand adaptability across a wide array of tasks and domains. This exposes a critical research gap: \textbf{How can an agent, relying solely on local information, dynamically make ``next hop'' routing decisions that contribute to an emergent, globally efficient solution across diverse tasks?}

To address this challenge, we situate our work within the framework of a \textbf{Self-Organizing Multi-Agent System} (SO-MAS). We define this as a decentralized network where individual agents autonomously make task-forwarding decisions using only local information, obviating the need for a central scheduler. In such a paradigm, we introduce \textit{BiRouter}, a novel routing method that empowers each agent to dynamically plan task pathways at runtime. Inspired by A* search algorithm \cite{AStar}, BiRouter makes intelligent, localized routing decisions through a bi-criteria heuristic that evaluates potential next-hop agents. This heuristic balances two metrics: (1) an \textbf{Importance (Imp) score}, which serves as a long-term heuristic by quantifying an agent's capability relevance to the overall task goal (analogous to A*'s heuristic cost, $h(n)$), and (2) a \textbf{Gap score}, which assesses an agent's contextual continuity to bridge the gap between the task's current state and the next logical step (akin to A*'s path cost, $g(n)$). By integrating these scores, BiRouter enables the emergence of globally coherent problem-solving pathways from simple, local rules. To enhance robustness in open, untrusted environments, the model further incorporates an iteratively updated agent reputation score into its decision-making process. Furthermore, to train our scoring functions and ensure their generalization, we constructed a large-scale, domain-rich dataset comprising thousands of annotated task-routing paths for diverse multi-step problems. Our main contributions can be summarized as follows:

\begin{itemize}
    \item We introduce BiRouter, a novel bi-criteria routing method for dynamic and decentralized task planning that balances long-term relevance with short-term state-awareness.
    \item We develop and release a large-scale and diverse dataset specifically designed for training multi-agent collaboration models, providing a valuable resource for the research community.
    \item We provide extensive experimental validation demonstrating that BiRouter significantly improves collaboration efficiency and robustness in both centralized and distributed settings, outperforming established baselines.
\end{itemize}

\section{Related Work}

\paragraph{LLM-driven Multi-Agent Systems.} 

The rise of LLMs has enabled Multi-Agent Systems (MAS) that outperform single agents on complex tasks \cite{Camel,AutoAgents,LLM-Debate,MADebate}. Early closed systems relied on predefined roles \cite{AgentVerse,AutoGen} or aggregating outputs \cite{LLM-Blender,MoA}, inherently limiting extensibility. To address this, the community has shifted towards open networks, developing foundational protocols like A2A \cite{A2A}, ANP \cite{ANP} and Agora \cite{Agora} for dynamic discovery and decentralized communication. While these protocols establish the infrastructure for interoperability, they primarily solve for connectivity, leaving a critical gap in the task-oriented, intelligent collaboration strategies needed for effective coordination.

\paragraph{Agent Collaboration Planning Mechanisms.}
Effective agent collaboration is crucial for task success. A dominant paradigm is static planning, which pre-computes a fixed workflow before task execution \cite{SPV,GPTSwarm}. Methods include pruning connections (AgentPrune \cite{AgentPrune}), generating task-specific structures (G-Designer \cite{G-Designer}, AFlow \cite{AFlow}), or determining a dependency-based order (MacNet \cite{MacNet}). However, these static approaches lack adaptability to dynamic changes or failures. To address this rigidity, dynamic approaches like DyLAN \cite{DyLAN} and MaAS \cite{MaAS} adapt team structures or select executors during the task. Despite their flexibility, both static and dynamic methods typically rely on a centralized planner with a global view of the system. This centralized model is not applicable in large-scale, decentralized environments where information is inherently local. Our work addresses this gap by proposing a mechanism for agents to make effective runtime decisions using only local information.

\paragraph{Heuristic Search Methods.} 
Planning a collaborative workflow can be framed as a pathfinding problem in a complex state space, making heuristic search highly relevant \cite{LLM-AStar,HeuNavi}. Recent methods have leveraged this, such as Tree-of-Thought \cite{ToT} using tree search for reasoning paths, and ToolChain* \cite{ToolChain} and AFlow applying A* \cite{AStar} and MCTS \cite{MCTS} to optimize workflows. However, A*'s reliance on a static heuristic makes it ill-suited for dynamic environments, while the more adaptable MCTS incurs significant computational overhead from extensive simulations. Inspired by the goal-oriented nature of A*, we introduce BiRouter: a novel, hybrid bi-criteria routing method. It navigates these dynamic spaces by learning its own heuristic functions, enabling efficient and adaptive task routing.

\section{Methodology}

\begin{figure*}[t!]
    \centering
    \includegraphics[width=\textwidth]{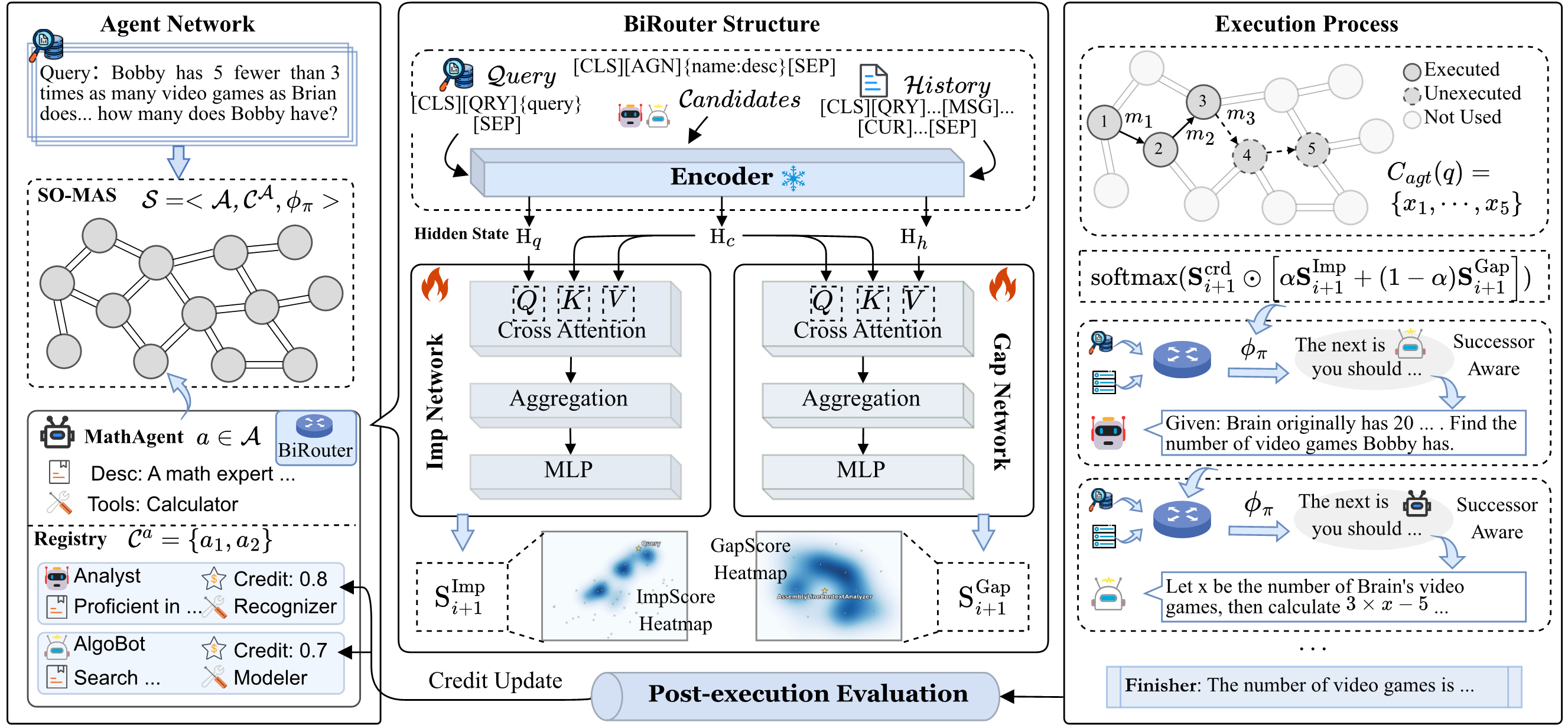}
    \caption{The architecture of BiRouter and its workflow in self-organizing agent systems.}
    \label{fig:overall_architecture}
\end{figure*}

Figure~\ref{fig:overall_architecture} presents the architecture of our method. Our system operates as a self-organizing agent network where any agent can initiate a query. It then employs its built-in routing mechanism, \textit{BiRouter}, to dynamically forward the task. This distributed routing process progressively approaches a optimal execution path, forming a runtime adaptive agent chain. To enhance system robustness, we employ a reputation mechanism that dynamically updates each agent's credit score upon task completion. In the following sections, Section~\ref{sec:preliminary} introduces definitions and optimization objectives, Section~\ref{sec:criteria} details the forwarding criteria, Section~\ref{sec:policy} details how selection policy of BiRouter helps to generate agent chain, and Section~\ref{sec:datagen} describes the generation strategy of our BiRouter training dataset.

\subsection{Preliminaries}
\label{sec:preliminary}

\paragraph{System and Agent Chain Definition.}
We define a Self-organizing Multi-agent System $\mathcal{S}$ as a decentralized network operating without a central coordinator or hierarchical structure:
\begin{equation}
\mathcal{S} = \left \langle \mathcal{A}, \mathcal{C}^{\mathcal{A}}, \phi_\pi \right \rangle  
\end{equation}
where $\mathcal{A} = \{a_i| i=1, \dots, n\}$ is the set of agents, each with a description $\mathbf{Desc}(a_i)$; $\mathcal{C}^{\mathcal{A}} = \{\mathcal{C}^{a_i} | a_i \in \mathcal{A}\}$ defines the set of known successor candidates for each agent $a_i$; and $\phi_\pi$ is the successor selection policy. In this system, each agent can partially execute a query and autonomously forward the task based on its local policy.

Given a query $q$, the system's objective is to form an \textit{agent chain}, $C_{agt}(q) = \{ x_1, x_2, \dots , x_L \}$, which is an ordered sequence of agents that collectively solve the task. The formation of this chain, denoted as $C_{agt} = \Pi(\mathcal{A},q)$, results from this decentralized decision-making process.

\paragraph{Problem Formulation.}
In contrast to centralized approaches that pre-plan the entire agent chain based on global knowledge, our system operates without a central scheduler. The agent chain emerges dynamically from a sequence of local, autonomous decisions. Each agent $x_i$ makes its decision based on a strictly local observation $o^{x_i}(q)$, which represents the agent's limited, egocentric view of the system:
\begin{equation}
o^{x_i} (q) = \overbrace{(\mathcal{H}_{i-1}^{(q)})}^{\text{Past}} \ \cup \ \overbrace{\mathbf{Desc}(x_i)}^{\text{Present}} \ \cup \ \overbrace{\{\mathbf{Desc}(x_j) | x_j \in \mathcal{C}^{x_i}\}}^{\text{Future}}
\end{equation}
where $\mathcal{H}_{i-1}^{(q)} = (q, x_1, m_1, \dots, x_{i-1}, m_{i-1})$ is the chain's history up to step $i-1$, and $m$ is the runtime message.

Our objective is to learn an optimal local policy $\pi^*$ that selects the best successor $c^{x_i*}$ from the local candidate set $\mathcal{C}^{x_i}$, rather than the global set $\mathcal{A}$:
\begin{equation}
c^{x_i*} = \phi_{\pi^*} (\mathcal{C}^{x_i} | o^{x_i}(q))
\end{equation}
The optimal policy $\pi^*$ favors successors based on a local decision score $\mathbf{S}^{x_i}$:
\begin{equation}
\pi^*(\mathcal{C}^{x_i} | o^{x_i}(q)) \propto \mathbf{S}^{x_i}
\end{equation}
This single-hop forwarding mechanism allows each agent to make the best possible local decision, progressively constructing a globally effective agent chain $C_{agt}^*$.

\subsection{Forwarding Criteria}
\label{sec:criteria}
The forwarding decision is governed by three key criteria:

\paragraph{Importance (ImpScore).} 
This score quantifies an agent's capability relevance to the overall task query. It is computed by a function $\operatorname{ImpScore}: \mathcal{A} \times \mathcal{Q} \to \mathbb{R}^{|\mathcal{A}| \times |\mathcal{Q}|}$, which outputs a matrix where each element $(i, j)$ represents the importance of agent $a_i$ for query $q_j$, where $\mathcal{Q}$ is the query set and $q_j\in\mathcal{Q}$. The score prioritizes agents with higher task-criticality, thereby enhancing execution efficiency.

\paragraph{Cohesion (GapScore).} 
This score evaluates how well a candidate agent fits into the emerging chain, ensuring logical and contextual continuity. The function $\operatorname{GapScore}: \mathcal{C}^{x_i} \to \mathbb{R}^{|\mathcal{C}^{x_i}|}$ computes a vector of compatibility scores for the set of candidate successors $\mathcal{C}^{x_i}$ of the current agent $x_i$. A higher score indicates a smoother, more logical transition.

The heatmaps in Figure~\ref{fig:overall_architecture} show the semantic distribution of ImpScore and GapScore, where darker colors indicate higher scores. Agents with a high ImpScore form a path from the query to the goal, while those with a high GapScore form a cluster around the current agent.

\paragraph{Agent Reputation.} 
To ensure system reliability, the policy incorporates agent reputation, a dynamic value represented by a credit score $\mathbf{S}^{\text{crd}} \in \mathbb{R}^{|\mathcal{A}|}$, reflecting each agent's historical trustworthiness and performance. This score is updated post-execution, where an LLM evaluates the agent's performance to determine a multiplicative update factor.

\subsection{The Successor Selection Policy}
\label{sec:policy}

At each step $i$, the current agent $x_i$ selects a successor $x_{i+1}$ from its candidates $\mathcal{C}^{x_i}$. This decision is executed by its built-in \textbf{BiRouter} module, which implements our policy $\phi_\pi$. The BiRouter consists of two branches that process inputs through a shared encoder, followed by distinct Cross Attention and MLP layers to compute the ImpScore and GapScore, respectively:
\begin{itemize}
    \item \textbf{Importance Branch} ($\mathcal{R}_{\text{Imp}}$): Computes $\mathbf{S}^{\text{Imp}}$, assessing the semantic relevance of each candidate to the original query $q$.
    \item \textbf{Cohesion Branch} ($\mathcal{R}_{\text{Gap}}$): Computes $\mathbf{S}^{\text{Gap}}$, evaluating the contextual fit of each candidate as a successor to $x_i$.
\end{itemize}
Formally, the score vectors for the next step ($i+1$) are computed as:
\begin{equation}
\begin{split}
    \mathbf{S}_{i+1}^{\text{Imp}}(q) &= \mathcal{R}_{\text{Imp}}(q, \mathbf{Desc}(\mathcal{C}^{x_i})) \\
    \mathbf{S}_{i+1}^{\text{Gap}}(q) &= \mathcal{R}_{\text{Gap}}(o^{x_i}(q))
\end{split}
\label{eq:scores}
\end{equation}
where $\mathcal{R}_{\text{Imp}}(\cdot)$ and $\mathcal{R}_{\text{Gap}}(\cdot)$ are the neural network functions of the BiRouter, and the outputs $\mathbf{S}_{i+1}^{\text{Imp}}, \mathbf{S}_{i+1}^{\text{Gap}} \in \mathbb{R}^{|\mathcal{C}^{x_i}|}$.

\paragraph{Probabilistic Synthesis.}
These dynamic scores are integrated with the static agent reputation scores ($\mathbf{S}^{\text{crd}}$) to produce a final probability distribution for successor selection. The local policy $\pi_i$ is defined as:
\begin{equation}
\begin{split}
    & \mathcal{L}ogits_{i+1} = \mathbf{S}^{\text{crd}}_{i+1} \odot \left[ \alpha \mathbf{S}_{i+1}^{\text{Imp}} + (1-\alpha)\mathbf{S}_{i+1}^{\text{Gap}} \right] \\
    &\pi_i(\mathcal{C}^{x_{i}} | x_i, q) = \text{Softmax}(\mathcal{L}ogits_{i+1})
\end{split}
\end{equation}
where $\mathbf{S}^{\text{crd}}_{i+1}$ is the vector of credit scores for the candidates, $\odot$ denotes the element-wise product, and $\alpha \in [0,1]$ is a hyperparameter balancing importance and cohesion. This formulation allows an agent's reputation to act as a multiplicative gate on the combined task-relevance scores.

A key feature of our design is that the transition message $m_i$ is generated \textbf{after} the successor $x_{i+1}$ has been selected. This enables the current agent $x_i$ to craft a \textbf{successor-aware message}, tailoring its content to the specific capabilities of $x_{i+1}$ to ensure a seamless and efficient task handoff:
\begin{equation}
m_i = x_i(\mathcal{H}_{i-1}^{(q)}, \mathbf{Desc}(x_{i+1}))
\end{equation}

\paragraph{Overall Chain Probability.}
The formation of a complete agent chain $C_{agt}(q)$ is initiated from a user-specified starting agent $x_1$. The probability of a specific chain, given the starting agent, is the product of the individual transition probabilities determined by our local policy $\pi$:
\begin{equation}
  \mathcal{P}(C_{agt} | x_1, q) = \prod_{i=1}^{L-1} \pi(\mathcal{C}^{x_{i}}| x_{i}, q)
  \label{eq:chain_prob}
\end{equation}

\paragraph{Adaptive Task Termination.}
We introduce a special \textbf{Finisher} agent to enable adaptive task completion. As a potential candidate at each step, its selection concludes tasks based on the task state rather than a fixed length, thus reducing unnecessary token consumption.
\begin{figure}
    \centering
    \includegraphics[width=1\linewidth]{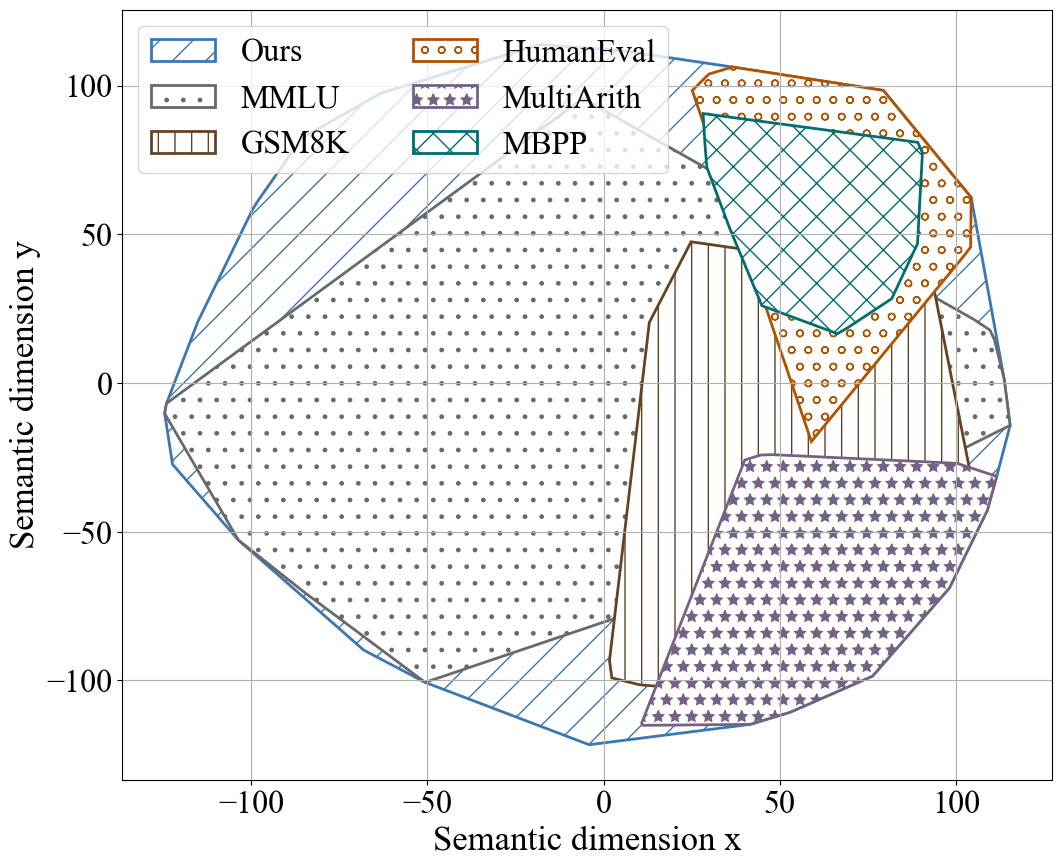}
    \caption{Distribution of MARS against others in a 2D semantic space projected by t-SNE.}
    \label{fig:dataset_distribution}
\end{figure}

\subsection{Training Data Generation}
\label{sec:datagen}
To foster the BiRouter's cross-domain generalization, we constructed Multi-agent Routing Dataset (MARS), a diverse dataset for BiRouter training. MARS consists of two main components: $\mathcal{D}_I = \{(q^{(i)}, a^{(i)}; s^{(i)}_I)\}$ for ImpScore, which pairs queries and agents with importance scores, and $\mathcal{D}_G = \{(c^{(j)}, a^{(j)}; s^{(j)}_G)\}$ for GapScore, which pairs execution contexts and agents with cohesion scores.

We initiated by prompting an LLM to produce a broad set of domain keywords, which were manually curated into 115 representative domains. Then, we prompted the LLM to generate unique queries within one domain or cross domains. To ensure comprehensive semantic coverage, we densified the query embedding space by identifying and populating sparse regions. This process was guided by a semantic density function using an RBF kernel:
\begin{equation}
\rho(x) = \frac{1}{N_q} \sum_{i=1}^{N_q} \exp\left(-\frac{\|x - q_i\|^2}{2\sigma^2}\right)
\end{equation}
where the LLM was prompted to generate new queries in regions of low density $\rho(x)$.

For each query, an LLM generated multiple candidate execution paths (i.e., agent sequences). Within each path, the LLM ranked agents by their criticality; these ranks were averaged across multiple iterations for robustness. To promote efficiency, a penalty factor $\gamma$ was applied to agents in long execution paths known for high token consumption. The final ImpScore is calculated using a scaled Sigmoid function to amplify the distinction between critical and non-critical agents within a target interval $[l, u]$:
\begin{equation}
s_I(a_i) = [l + (u-l) \cdot \frac{1}{1 + e^{-\beta(N_r-r_i)}}] \cdot \gamma
\end{equation}
where $N_r$ is the number of agents, $r_i$ is the agent's rank, $l$ and $u$ are set to 0.3 and 1, and $\beta=2$ is a scaling parameter.

\begin{table*}[t]
\centering
\small
\begin{tblr}{
  colspec = {lccccccccc},
}
\toprule
Method & Dyn. & Dis. & MMLU & GSM8K & SVAMP & HumanEval & MultiArith & MBPP & Avg. \\
\midrule
Single-agent & -- & -- & 77.81 & 87.45 & 88.26 & 87.08 & 96.85 & 71.83 & 84.88 \\
CoT & -- & -- & 78.43$_{+0.62}$ & 87.10$_{-0.35}$ & 86.24$_{-2.02}$ & 88.13$_{+1.05}$ & 96.31$_{-0.54}$ & 71.83$_{+0.00}$ & 84.67 \\
ComplexCoT & -- & -- & 81.05$_{+3.24}$ & 86.89$_{-0.56}$ & 90.53$_{+2.27}$ & 87.49$_{+0.41}$ & 96.70$_{-0.15}$ & 72.36$_{+0.53}$ & 85.84 \\
SC (CoT$\times$5) & -- & -- & 80.96$_{+3.15}$ & 87.57$_{+0.12}$ & 87.92$_{-0.34}$ & 88.60$_{+1.52}$ & 96.58$_{-0.27}$ & 73.60$_{+1.77}$ & 85.87 \\
\midrule
MacNet & $\times$ & $\times$ & 82.98$_{+5.17}$ & 87.95$_{+0.50}$ & 88.06$_{-0.20}$ & 84.57$_{-2.51}$ & 96.03$_{-0.82}$ & 65.28$_{-6.55}$ & 84.15 \\
LLM-Blender & $\times$ & $\times$ & 81.22$_{+3.41}$ & 88.35$_{+0.90}$ & 89.52$_{+1.26}$ & 88.80$_{+1.72}$ & 97.29$_{+0.44}$ & 77.05$_{+5.22}$ & 87.04 \\
LLM-Debate & $\times$ & $\times$ & 81.04$_{+3.23}$ & 89.47$_{+2.02}$ & \underline{91.76}$_{+3.50}$ & 88.68$_{+1.60}$ & 97.33$_{+0.48}$ & 70.29$_{-1.54}$ & 86.43 \\
GPTSwarm & $\times$ & $\times$ & 82.80$_{+4.99}$ & 89.14$_{+1.69}$ & 87.02$_{-1.24}$ & 89.32$_{+2.44}$ & 96.79$_{-0.06}$ & 77.43$_{+5.60}$ & 87.08 \\
G-Designer & $\times$ & $\times$ & \textbf{87.20}$_{+9.39}$ & \underline{93.97}$_{+6.52}$ & 90.29$_{+2.03}$ & 87.50$_{+0.42}$ & 98.33$_{+1.48}$ & -- & -- \\
\midrule
AgentVerse & $\checkmark$ & $\times$ & 78.36$_{+0.55}$ & 89.91$_{+2.46}$ & 89.64$_{+1.38}$ & 89.29$_{+2.21}$ & 97.50$_{+0.65}$ & 74.28$_{+2.45}$ & 86.50 \\
DyLAN & $\checkmark$ & $\times$ & 79.96$_{+2.15}$ & 89.98$_{+2.53}$ & 88.48$_{+0.22}$ & 90.42$_{+3.34}$ & 97.12$_{+0.27}$ & 77.30$_{+5.47}$ & 87.21 \\
MaAS & $\checkmark$ & $\times$ & -- & 92.30$_{+4.85}$ & -- & \textbf{92.85}$_{+5.77}$ & \underline{98.80}$_{+1.95}$ & \underline{82.17}$_{+10.34}$ & -- \\
\midrule
\textbf{BiRouter (Ours)} & $\checkmark$ & $\checkmark$ & \underline{86.80}$_{+8.99}$ & \textbf{94.09}$_{+6.64}$ & \textbf{93.20}$_{+4.04}$ & \underline{91.46}$_{+4.38}$ & \textbf{100.00}$_{+3.15}$ & \textbf{84.82}$_{+12.99}$ & \textbf{91.73} \\
\bottomrule
\end{tblr}
\captionsetup{format=plain, justification=justified, singlelinecheck=false}
\caption{Performance comparison with BiRouter and baselines in centralized MAS settings. ``Dyn.'' and ``Dis.'' indicate whether a method supports dynamic topology and distributed execution ($\checkmark$: yes, $\times$: no).}
\label{tab:final_results}
\end{table*}

The context for GapScore is derived from the generated agent sequences. An agent $a_i$ at step $k$ receives a GapScore based on its position in the full path:
\begin{equation}
s_G(a_i, k) = 
\begin{cases}
    1 / (i-k+1) & \text{if } i \geq k \\
    0 & \text{if } i < k 
\end{cases}
\end{equation}
where $i$ is the agent's index.
Finally, a dedicated ``Finisher'' agent was appended to each path to signal task completion. As shown in Figure~\ref{fig:dataset_distribution}, a t-SNE visualization of the dataset reveals a rich semantic distribution, which is crucial for enhancing the BiRouter's cross-domain generalization.

\begin{figure}[t!]
    \centering
    \includegraphics[width=1.0\linewidth]{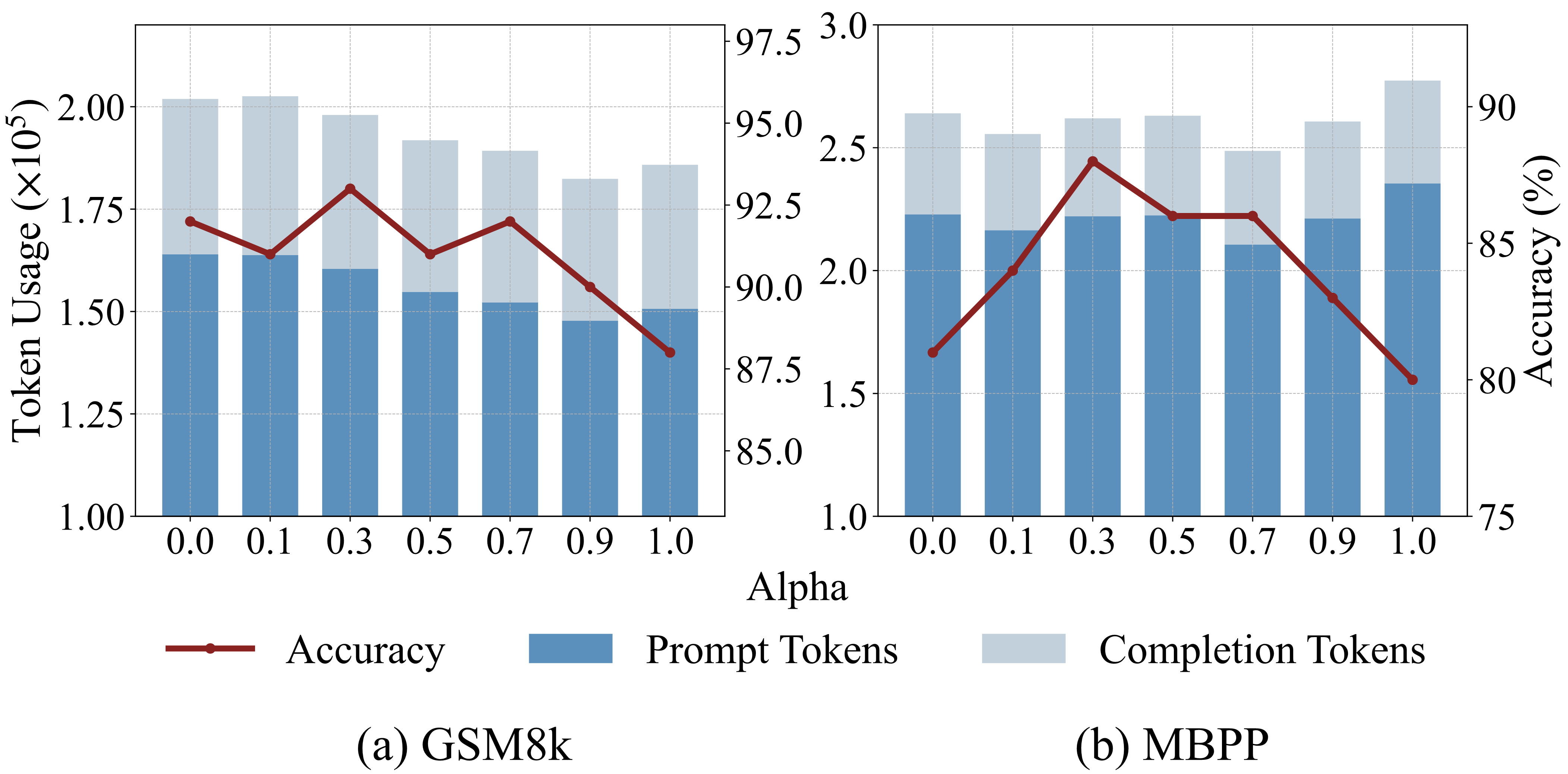}
    \captionsetup{format=plain, justification=justified, singlelinecheck=false}
    \caption{BiRouter's performance and token consumption on GSM8K and MBPP across different $\alpha$ values.}
    \label{fig:alpha_graph}
\end{figure}

\section{Experiments}

\subsection{Experimental Setup}

\paragraph{Benchmarks.}

We evaluate BiRouter on six public benchmarks: MMLU \cite{MMLU}, GSM8K \cite{GSM8K}, MultiArith \cite{MultiArith}, SVAMP \cite{SVAMP}, HumanEval \cite{HumanEval} and MBPP \cite{MBPP}.

\paragraph{Baselines.}
We compare BiRouter with three series of agentic baselines: (1) \textbf{Single-agent approach}: Single-agent LLM, CoT \cite{CoT}, ComplexCoT \cite{ComCoT}, and CoT-SC \cite{SC-CoT}. (2) \textbf{Static coordination MAS}: MacNet \cite{MacNet}, LLM-Blender \cite{LLM-Blender}, LLM-Debate \cite{LLM-Debate}, GPTSwarm \cite{GPTSwarm}, G-Designer \cite{G-Designer}. (3) \textbf{ Dynamic coordination MAS}: AgentVerse \cite{AgentVerse}, DyLAN \cite{DyLAN}, MaAS \cite{MaAS}.

\paragraph{Implementation Details.}
We use \texttt{qwen3-embedding-0.6b} as the frozen encoder for BiRouter, training only the Cross-Attention and scoring head on our synthetically generated data. All agents use \texttt{gpt-4o-mini} via the OpenAI API. We set the \texttt{temperature} to 0 for all single-agent methods and 1 for all multi-agent methods. We use \texttt{gpt-4.1} to generate agent profiles to build the agent network, including agents specialized for the benchmarks and one fixed ``Finisher'' agent for task exit and answer summarization. We conduct each experiment twice and take the average as the final result. The optimal hyperparameter $\alpha$ was tuned using $Q \in \{100, 200\}$ queries randomly sampled from GSM8K and MBPP. As illustrated in Figure~\ref{fig:alpha_graph}, we evaluated BiRouter's performance across various $\alpha$ values. Balancing performance against token consumption, we set the final value to $\alpha=0.3$.

\subsection{Results in Centralized MAS}

We first conducted experiments in a centralized MAS environment where all agents are available at each stage of the task. We compared our method with 12 other baselines, with all agents powered by \texttt{gpt-4o-mini}. The evaluation results demonstrate that BiRouter is:

\paragraph{High-performing.} As summarized in Table~\ref{tab:final_results}, BiRouter achieves state-of-the-art or highly competitive results across all benchmarks. The average accuracy confirms that BiRouter surpasses all other baseline categories, demonstrating performance gains of 5.86\% to 7.06\% over single-agent methods, 4.65\% to 7.58\% over static coordination methods, and 4.52\% to 5.23\% over dynamic coordination methods. BiRouter achieved optimal results on 4 out of 6 benchmarks, demonstrating strong task performance.

\begin{figure}[t!]
    \centering
    \includegraphics[width=1.0\linewidth]{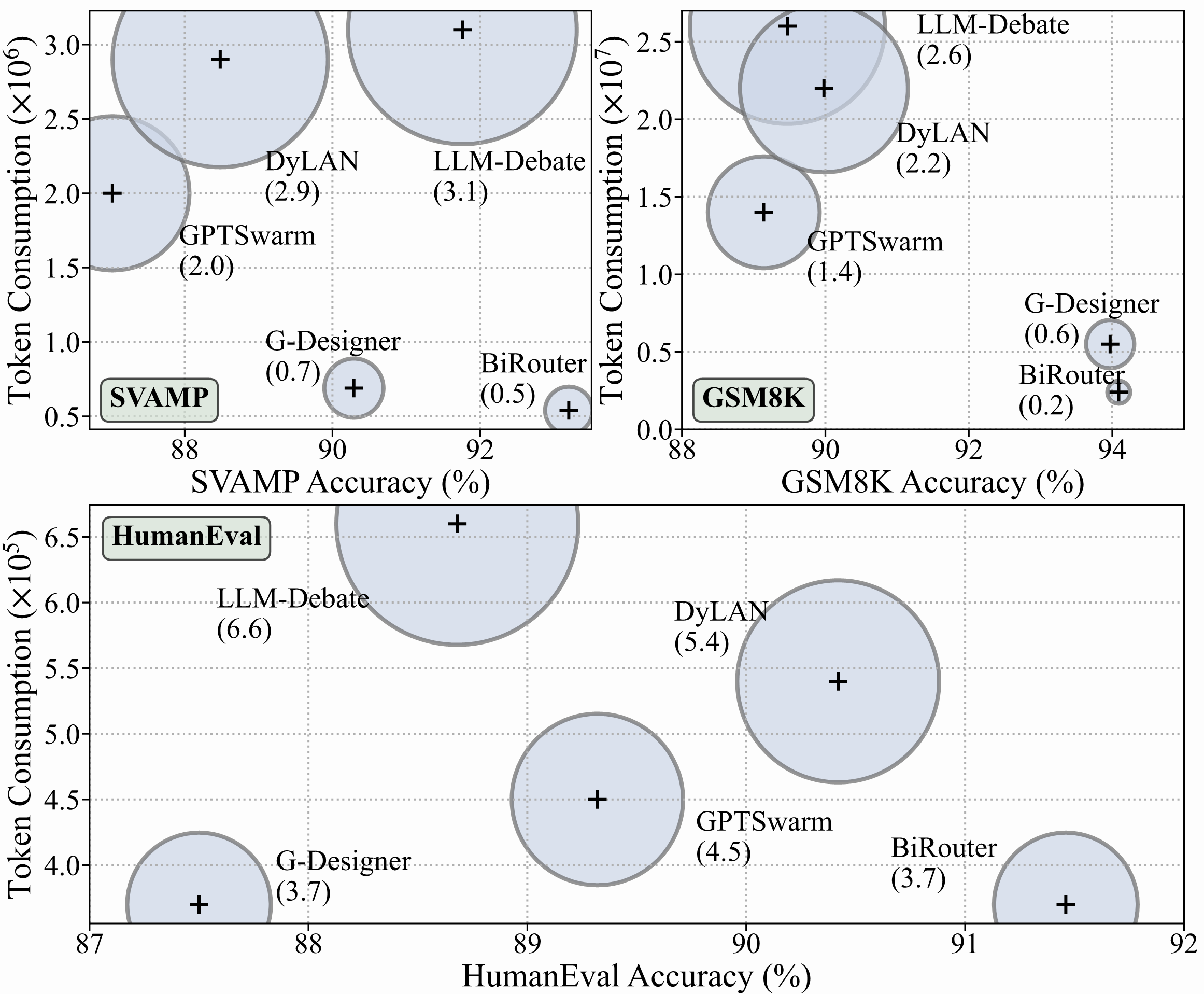}
    \captionsetup{format=plain, justification=justified, singlelinecheck=false}
    \caption{Visualization of the performance metrics and token consumption. The diameter of each point is proportional to its $y$-axis value.}
    \label{fig:token_cost}
\end{figure}

\begin{table}[t!]
\centering
\setlength{\tabcolsep}{3pt} 
\begin{tabular*}{\columnwidth}{@{\extracolsep{\fill}} l|cc|cc}
\toprule
\multirow{3}{*}{\textbf{Method}} & \multicolumn{2}{c}{\textbf{GSM8K}} & \multicolumn{2}{c}{\textbf{HumanEval}} \\
\cmidrule(lr){2-3} \cmidrule(lr){4-5}
& \textbf{Acc.}(\%) & \textbf{Cost} & \textbf{Pass@1}(\%) & \textbf{Cost} \\
\midrule
DyLAN & 87.95 & $6.3\times10^6$ & 68.29 & $6.2\times10^5$ \\
MaAS & 86.43 & $3.8\times10^6$ & 79.27 & $4.9\times10^5$ \\
\midrule
BiRouter & 91.99 & $2.8\times10^6$ & 89.63 & $4.6\times10^5$ \\
\bottomrule
\end{tabular*}
\captionsetup{format=plain, justification=justified, singlelinecheck=false}
\caption{Performance and token consumption comparison in SO-MAS settings. ``Acc.'' denotes accuracy, and ``Cost'' represents the number of tokens used.}
\label{tab:dis_results}
\end{table}

\paragraph{Token-economical.} Figure~\ref{fig:token_cost} visualizes the trade-off between token consumption (both prompt and completion tokens) and performance. Across both SVAMP and GSM8K, BiRouter not only surpasses all other baselines in performance but does so with the lowest token expenditure. On HumanEval, it delivers a 3.96\% performance boost over G-Designer while maintaining a comparable token count. This stands in sharp contrast to methods like GPTSwarm and DyLAN, whose complex, iterative approaches inflate token costs. By comparison, BiRouter exemplifies an exceptional balance between high performance and resource efficiency.

\paragraph{Generalizable.} The consistent high performance across diverse domains (general reasoning, math, and code) demonstrates BiRouter's strong generalization capabilities, especially given it was trained exclusively on synthetically generated data. Its task-adaptive and dynamically adaptive routing mechanism allows it to adjust agent selection in real-time, showing its scalability in dynamic environments.

\subsection{Results in SO-MAS}

In this section, we simulate a self-organizing MAS (SO-MAS) environment where each agent can only communicate with three other randomly selected agents and must choose its successor from this limited set. We compare BiRouter with two other dynamic methods capable of operating in such a setting, DyLAN and MaAS, with results shown in Table~\ref{tab:dis_results}. The localized information in this decentralized setting leads to a general performance degradation across all methods. However, BiRouter exhibits the smallest performance drop, achieving 91.99 on GSM8K and 89.63 on HumanEval, respectively. This resilience stems from its native design for local decision-making. In contrast, DyLAN and MaAS rely heavily on the availability of global information, leading to a significant performance decline. BiRouter's strong performance in both centralized and decentralized architectures underscores its cross-architecture adaptability.

\begin{figure}[t!]
    \centering
    \includegraphics[width=1\linewidth]{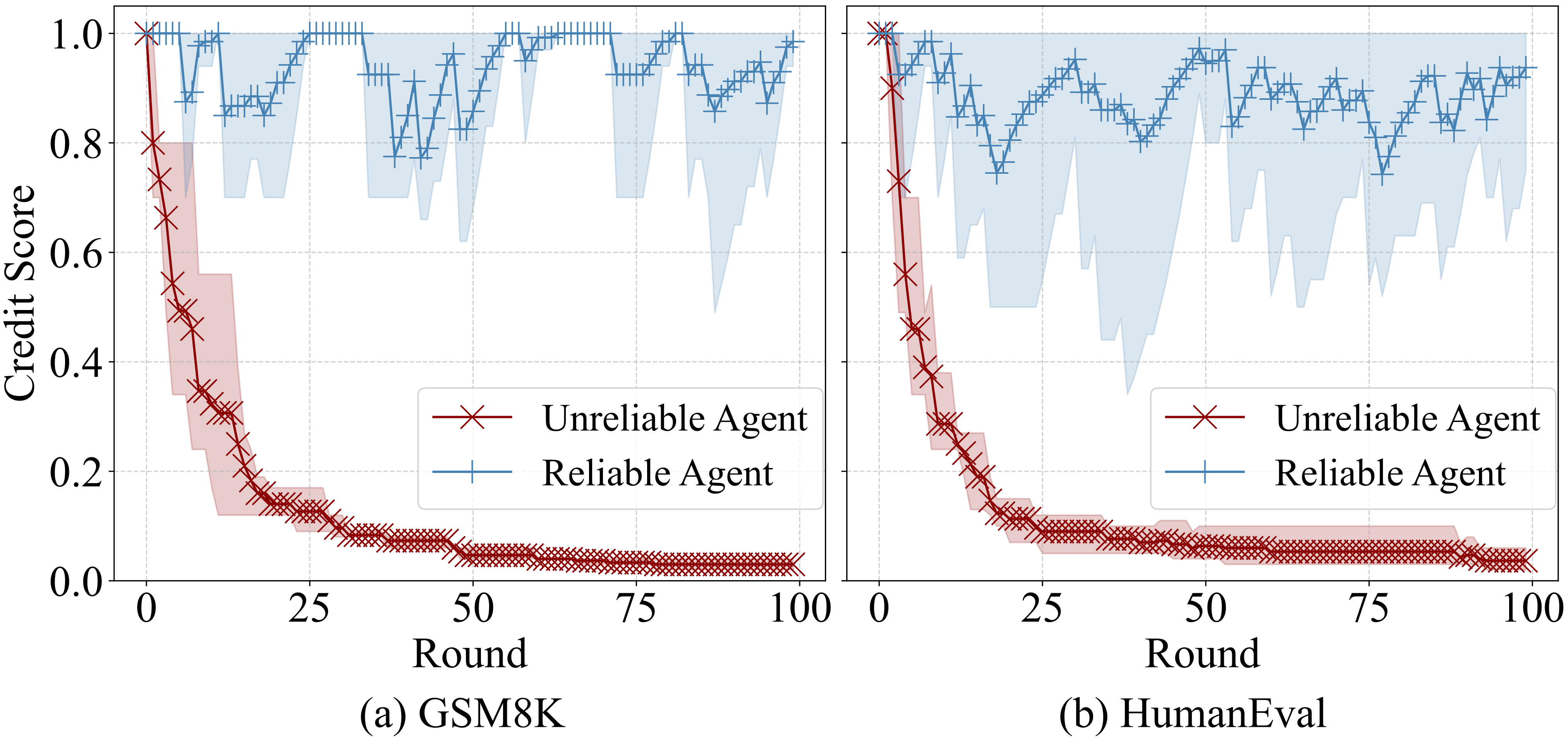}
    \captionsetup{format=plain, justification=justified, singlelinecheck=false}
    \caption{Agent credit score dynamics on GSM8K and HumanEval benchmarks.}
    \label{fig:credit}
\end{figure}
\begin{figure}[t!]
    \centering
    \includegraphics[width=1\linewidth]{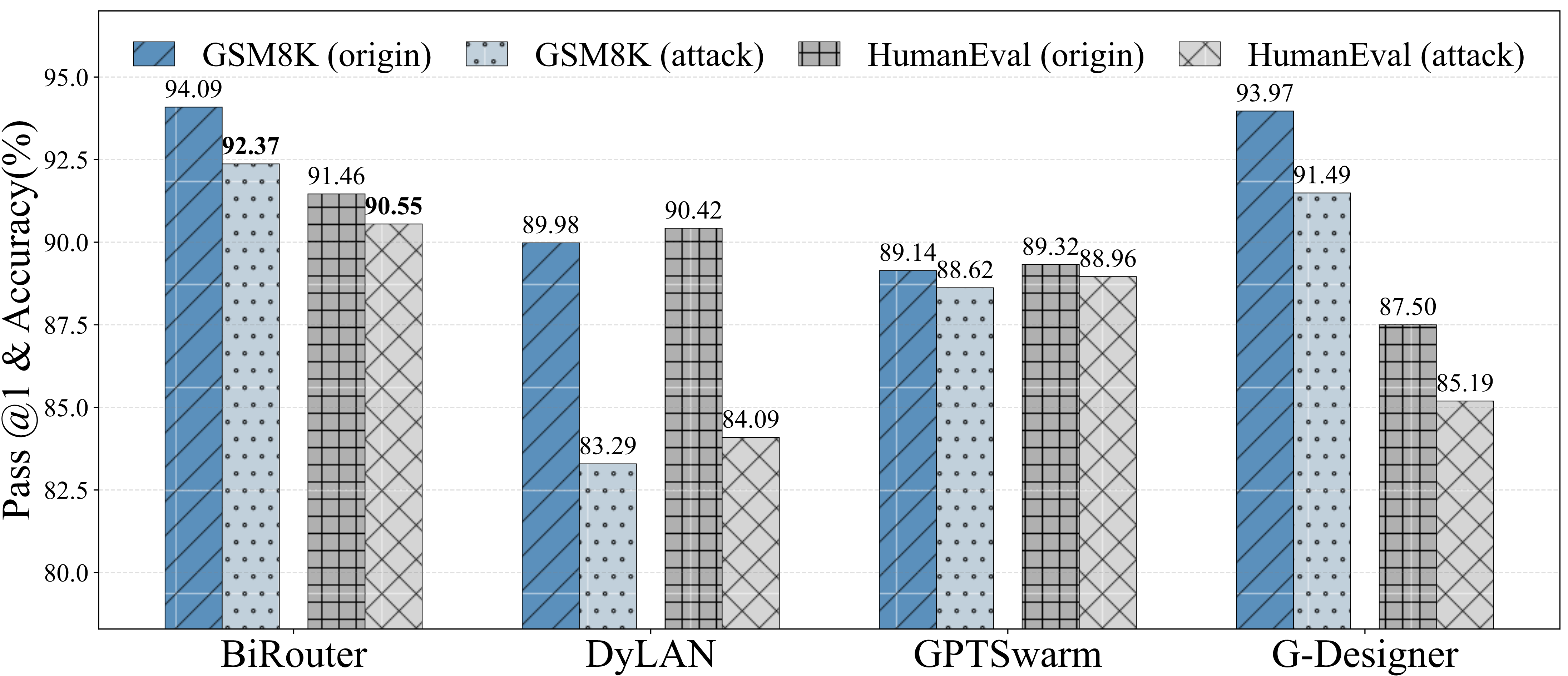}
    \captionsetup{format=plain, justification=justified, singlelinecheck=false}
    \caption{Performance comparison between before and after unreliable agent attack.}
    \label{fig:robustness}
\end{figure}

\subsection{Robustness}

To evaluate BiRouter's robustness, we introduced an equal number of unreliable agents into the network, created by using \texttt{gpt-4.1} to rewrite agent prompts to intentionally produce incorrect outputs. As shown in Figure~\ref{fig:credit}, which plots agent credit scores during experiments on GSM8K and HumanEval, BiRouter rapidly identifies and down-weights these unreliable agents, mitigating their negative influence. Consequently, as seen in Figure~\ref{fig:robustness}, while all methods suffer from performance degradation, BiRouter remains the highest performance under unreliable agent attacks. This demonstrates its robustness in open, untrustworthy environments.

\subsection{Case Study}

\begin{figure}[t!]
    \centering
    \includegraphics[width=1\linewidth]{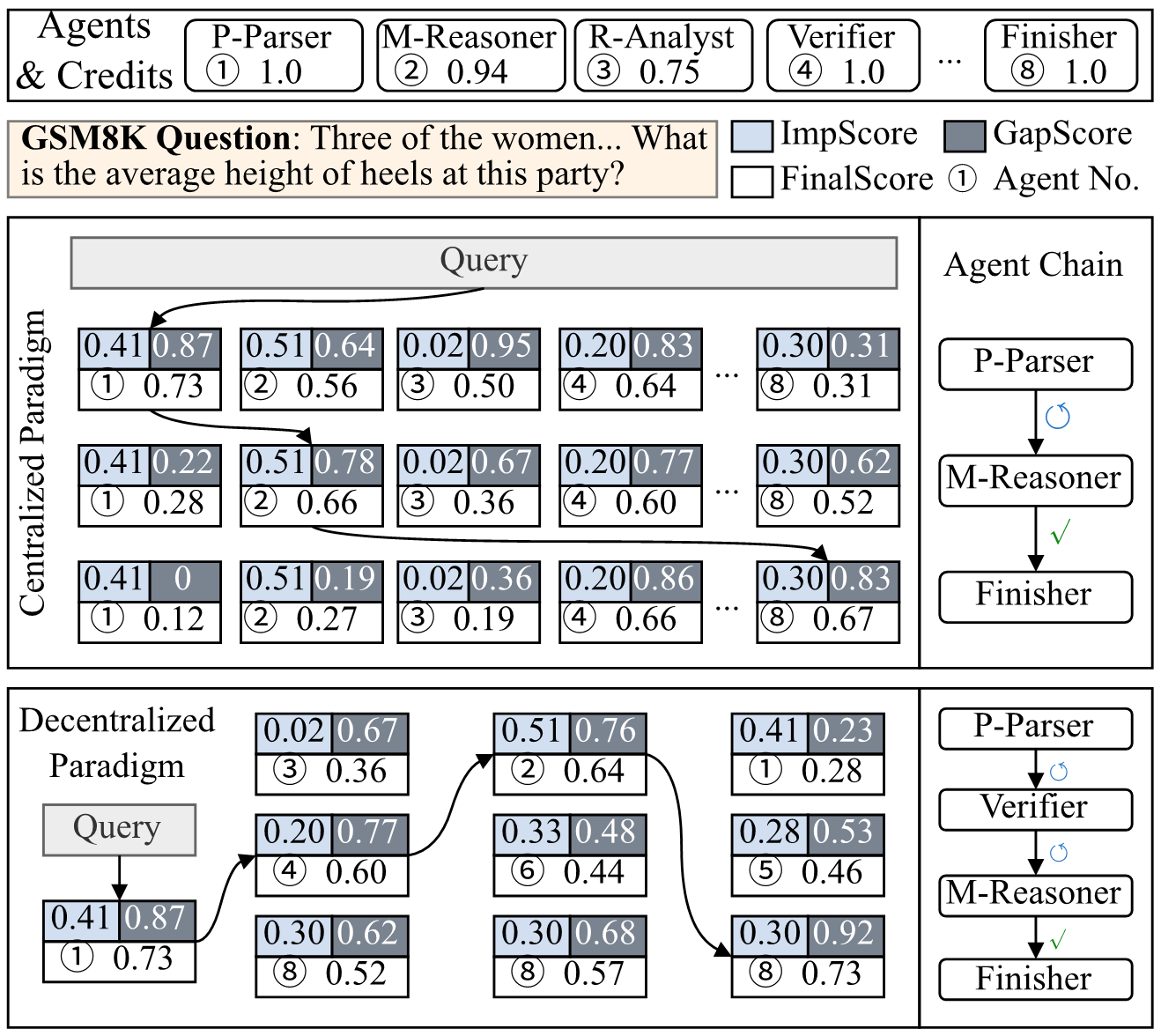}
    \captionsetup{format=plain, justification=justified, singlelinecheck=false}
    \caption{Visualization of BiRouter's routing mechanism across centralized and decentralized paradigms.}
    \label{fig:case_study_1}
\end{figure}
\begin{figure}[t!]
    \centering
    \includegraphics[width=1\linewidth]{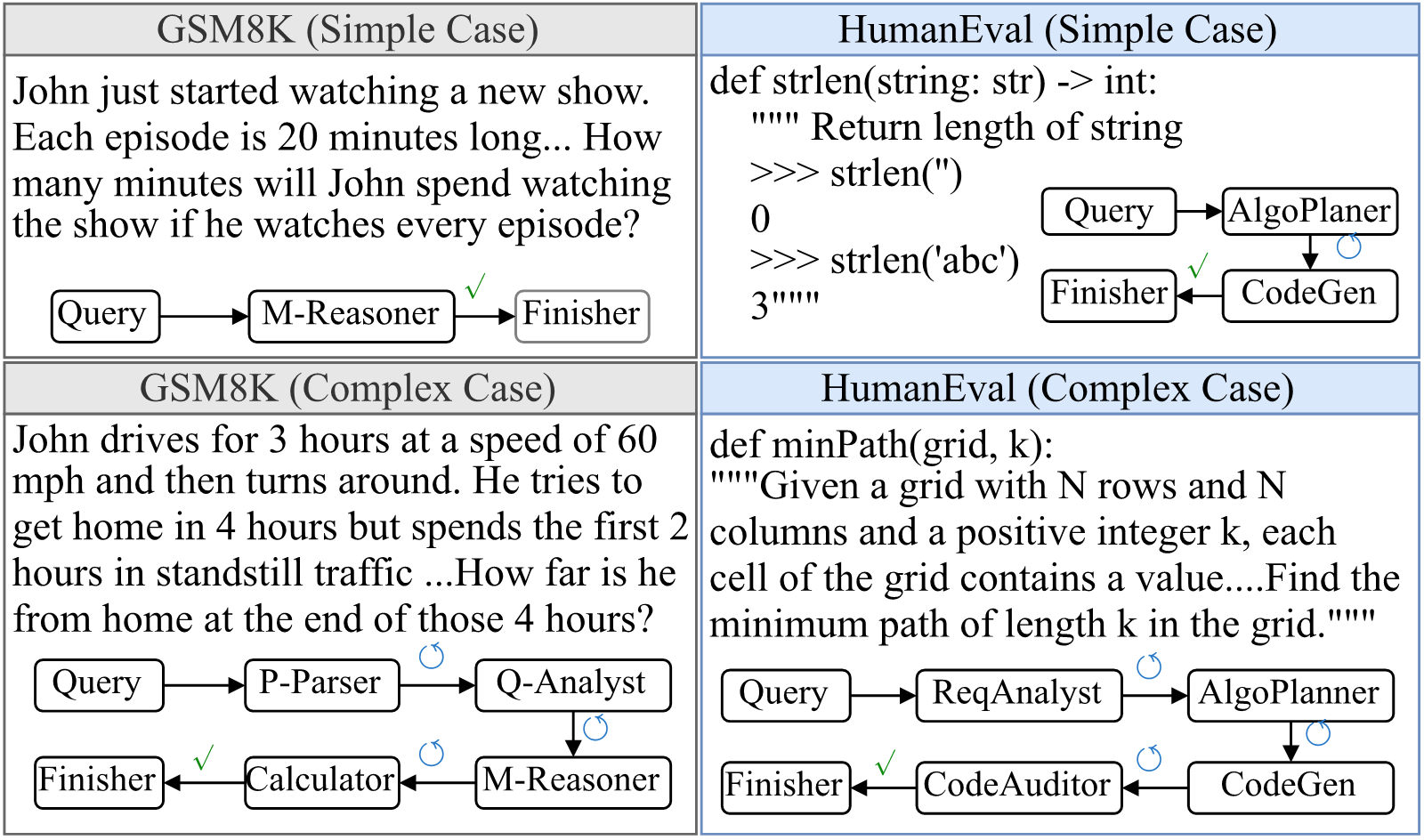}
    \captionsetup{format=plain, justification=justified, singlelinecheck=false}
    \caption{Case study on GSM8K and HumanEval tasks.}
    \label{fig:case_study_2}
\end{figure}

We present a case study on GSM8K and HumanEval to illustrate BiRouter's mechanism. As shown in Figure~\ref{fig:case_study_1}, in the centralized setting, BiRouter efficiently selects agents based on global visibility: ProblemParser (0.73) initiates, followed by MathReasoner (0.66) for reasoning and computation, and finally Finisher (the highest GapScore of 0.83) upon task completion. In the decentralized setting, with limited visibility, BiRouter initially selects the locally optimal Verifier (0.60) but corrects its path by routing to MathReasoner (0.64), still reaching the solution. Figure~\ref{fig:case_study_2} demonstrates BiRouter's flexibility across different tasks. For simple problems, it forms short chains of only one or two hops. For more complex tasks, it dynamically constructs longer chains by recruiting agents like QuestionAnalyst and CodeAuditor to aid in reasoning and decision-making, thereby improving the task success rate.

\subsection{Ablation Study}

We conducted ablation studies on GSM8K and HumanEval with the following configurations: (1) \textbf{w/o $\textbf{S}^{\text{crd}}$}: Agent credit scores are fixed at 1. (2) \textbf{w/o Succ.}: The current agent is not informed of the next-hop agent. (3) \textbf{w/o Fin.}: The Finisher agent is removed, and the number of execution steps is fixed at 5. The results in Tables~\ref{tab:final_ablation} show that all three components contribute significantly. Without the Reputation mechanism, BiRouter fails to recognize unreliable agents, degrading performance in untrustworthy networks. Removing the Successor-aware mechanism weakens inter-agent cooperation and lowers collaborative efficiency. The absence of the Finisher agent eliminates the ability for early task completion, forcing a fixed execution length. This reduces dynamic adaptability and leads to unnecessary token consumption, increasing the token overhead by approximately 69\% on GSM8K and 50\% on HumanEval.

\begin{table}[t!]
\centering
\small
\setlength{\tabcolsep}{2pt}
\begin{tabular}{l|cc|cc|cc|cc}
\toprule
& \multicolumn{4}{c|}{\textbf{GSM8K}} & \multicolumn{4}{c}{\textbf{HumanEval}} \\
\cmidrule(lr){2-5} \cmidrule(lr){6-9}
\textbf{Variant} & \multicolumn{2}{c|}{Origin} & \multicolumn{2}{c|}{Attack} & \multicolumn{2}{c|}{Origin} & \multicolumn{2}{c}{Attack} \\
& Acc. & Cost & Acc. & Cost & Pass & Cost & Pass & Cost \\
& (\%) & ($10^6$) & (\%) & ($10^6$) & (\%) & ($10^5$) & (\%) & ($10^5$) \\
\midrule
BiRouter & 94.09 & 2.1 & 92.37 & 2.5 & 91.46 & 3.5 & 90.55 & 3.7 \\
\midrule
$w/o$ $\textbf{S}^{\text{crd}}$ & 92.31 & 2.5 & 84.79 & 3.2 & 90.24 & 3.7 & 82.93 & 4.9 \\
$w/o$ Succ. & 90.84 & 2.5 & 90.35 & 3.0 & 87.80 & 3.8 & 87.20 & 3.9 \\
$w/o$ Fin. & 90.08 & 4.1 & 88.73 & 4.2 & 88.72 & 5.4 & 88.10 & 5.4 \\
\bottomrule
\end{tabular}
\captionsetup{format=plain, justification=justified, singlelinecheck=false}
\caption{Ablation study of BiRouter on GSM8K and HumanEval.}
\label{tab:final_ablation}
\end{table}

\section{Conclusion}

This paper addresses the critical challenge of task planning in open, decentralized multi-agent systems. We introduce BiRouter, a novel bi-criteria routing method designed for self-organizing MAS, which enables individual agents to make autonomous ``next-hop'' routing decisions based solely on local information. By balancing overall task relevance and contextual continuity of candidate agents at runtime, BiRouter dynamically constructs collaborative pathways, demonstrating high adaptability and scalability. Experiments show that BiRouter holds significant advantages in performance, token efficiency, and robustness, and can seamlessly adapt to both centralized and decentralized architectures. Our work provides an effective solution for achieving emergent intelligent collaboration. Future work will focus on online adaptive optimization and more complex collaboration patterns.

\bibliography{ref}

\end{document}